\documentclass{emulateapj}

\usepackage{natbib}

\usepackage{graphicx}

\renewcommand{\d}{\mathrm{d}}

\slugcomment{ApJ accepted}

\shorttitle{Comparisons of mass profiles in strong lensing}

\shortauthors{Shu et al.}

\begin{document}

\input epsf

\title{Comparisons between isothermal and NFW mass profiles for strong-lensing galaxy clusters}

\author
 {Chenggang Shu\altaffilmark{1,2,6} Binglu Zhou\altaffilmark{1,2}, Matthias
  Bartelmann\altaffilmark{2}, \\ Julia M. Comerford\altaffilmark{3}, J.-S. Huang\altaffilmark{4},
  Yannick Mellier\altaffilmark{5}}

\email{cgshu@shao.ac.cn}

\altaffiltext{1} {Shanghai Normal University, 100 Guilin Road,
Shanghai 200234, China}

\altaffiltext{2} {Zentrum f\"ur Astronomie der Universit\"at
Heidelberg, ITA, Albert-\"Uberle-Str. 2, 69120 Heidelberg, Germany}

\altaffiltext{3} {Astronomy Department, 601 Campbell Hall,
University of California, Berkeley, CA 94720, USA}

\altaffiltext{4} {Harvard-Smithsonian Center for Astrophysics, 60
Garden Street, Cambridge, MA 02138, USA}

\altaffiltext{5} {Institut d'Astrophysique de Paris, UMR 7095, 98bis
BD Arago, 75014 Paris, France}

\altaffiltext{6} {Shanghai Astronomical Observatory, Chinese Academy
of Sciences, Shanghai 200030, China}

\begin{abstract}

While both isothermal and NFW-based mass models for galaxy clusters
are widely adopted in strong-lensing studies, they cannot easily be
distinguished based solely on observed positions of arcs and
arclets. We compare the magnifications  predicted for giant arcs
obtained from isothermal and NFW profiles, taking axially-symmetric
and asymmetric mass distributions into account. We find that arc
magnifications can differ strongly between the two types of density
profiles even if the image morphology is well reproduced.
Magnifications by lenses with NFW density profiles are usually
larger than those for lenses with singular or nearly singular
isothermal density profiles, unless the latter have large cores.
Asymmetries play an important role. We illustrate our results with
the two well-studied clusters MS~2137 and A~370.  We confirm earlier
results showing that both isothermal and NFW mass models can very
well reproduce the observed arcs, radial arcs and other arclets.
While the mass model for MS~2137 is not very well constrained, the
two types of mass models produce strongly differing critical curves
and caustics for A~370. We find that the NFW mass profile is
preferred for A~370. We identify new arclet candidates in the field
of A~370. Redshift estimates allowed by the lens model are
consistent with results in the literature, where available. Three
newly found counter-images are suggested to arise from an active,
dust-enshrouded star-forming galaxy at $z\approx1.1$.

\end{abstract}


\section{Introduction}

Strong lensing by galaxy clusters, giving rise to strongly distorted
arc-like images, plays an important role in studies of the mass
distributions of galaxy clusters. It is highly non-linear, weakly
affected by baryonic physics, and the data reduction is
comparatively simple because it focuses on the positions of arcs and
arclets (e.g.~Leonard et al 2007; Limousin et al 2006; Halkola,
Seitz \& Pannella 2006; Zekser et al 2006; Covone et al 2006;
Comerford et al 2006; Broadhurst et al 2005; Gavazzi 2005; Gavazzi
et al 2003; Smith et al 2005; Kneib et al 2003; Kneib et al 1993;
Mellier, Fort \& Kneib 1993; Hammer 1991). Cosmological applications
of strong lensing range from the use of clusters as gravitational
telescopes (Frye et al 2007; Ofek et al 2006; Chary, Stern \&
Eisenhardt 2005; Kneib et al 2004; Ellis et al 2001) to cosmological
constraints based on arc statistics (Fedeli et al 2008; Hamana et al
2005; Li et al 2005; Meneghetti et al. 2005a,b; Puchwein et al 2005;
Dalal \& Holder, 2004; Wambsganss et al 2004; Bartelmann et al
1998). Currently, about 100 clusters are known to contain large
arcs, including those newly discovered by Bolton et al (2006) in the
Sloan Digital Sky Survey data.

One way of investigating cluster mass distributions with strong
lensing starts  from an assumed mass profile, which is commonly
taken to be the well-known singular or non-singular isothermal
sphere (SIS or NIS, respectively)
\begin{equation}
\label{eq:SIS}
  \rho_{\rm SIS}\propto \frac{1}{r^2}\quad\mbox{or}\quad
  \rho_{\rm NIS}\propto \frac{1}{r^2+r_\mathrm{c}^2}\;,
\end{equation}
where $r$ is the radius from the cluster centre and $r_\mathrm{c}$
the core radius.

Another mass model commonly used for strong lensing is based on the
high-resolution numerical simulations of dark-matter halos in the
$\Lambda$CDM framework by Navarro, Frenk \& White (1996, 1997;
hereafter NFW), who found the density profile
\begin{equation} \label{eq:NFW}
  \rho_{\rm NFW}\propto\left(\frac{r}{r_\mathrm{s}}\right)^{-1}\,
  \left[1+\left(\frac{r}{r_\mathrm{s}}\right)^2\right]^{-1}\;,
\end{equation}
parameterized by the scale radius $r_\mathrm{s}$. While simulations
and X-ray observations  of galaxy clusters both support the NFW
profile and applications to strong-lensing studies have been
successful (Zekser et al 2006; Comerford et al 2006; Kneib et al
2003), the SIS and NIS models can also reproduce observed arcs
remarkably well in many clusters (e.g.~Kneib et al 1993; Mellier et
al 1993; Miradal-Escud\'e 1995; Gavazzi et al 2003; Gavazzi 2005).

So far, conclusive evidence that the density profiles of real and
simulated galaxy  clusters agree with each other has not been found.
Unfortunately, it is virtually impossible to distinguish isothermal
and NFW density profiles based solely on the positions of observed
arcs because of the near-degeneracy between these two density
profiles close to the scale radius. If radial arcs or multiple
tangential arcs at different distances from the cluster centre are
available, the situation improves substantially because they probe
the density profile at different radii (Gavazzi et al 2003; Gavazzi
2005; Tu et al 2008). While the positions of arcs and arclets are
determined by the first derivative of the lensing potential, their
magnifications and distortions depend on its second derivatives.
Different models reproducing the image geometry can thus yield
substantially different image magnifications.

The scientific goal of this paper is to investigate how relative arc
and arclet magnifications can be used to break degeneracies between
different mass models for strongly lensing clusters when combined
with positional and photometric information.

Two well-studied clusters, A~370 and MS~2137, are used for
illustration. The code for fitting lens models is based on Comerford
et al (2006) and will be briefly described below. Wherever needed,
we adopt the ``concordance'' flat cosmology with
$\Omega_\mathrm{m}=0.3$, $\Omega_\Lambda=0.7$ and Hubble constant
$h=0.7$.

The paper is structured as follows. Starting from a one-component
lens model,  we compare in Sect.~2 the predicted magnifications of
the resulting arcs for the SIS, NIS, and NFW profiles, respectively,
starting from axial symmetry and taking deviations therefrom into
account. The results are easily extrapolated to lens systems with
multiple components since the distance between two lens components
is typically larger than their core or scale radii, $r_\mathrm{c}$
or $r_\mathrm{s}$. In Sect.~3, earlier work on MS~2137 and A~370 is
briefly summarized. We adapt different mass models to these
clusters, derive the arc magnifications and compare with
observations. Based on the high-quality images observed, we study
the arclets in the field of A~370 in detail. We present our
conclusions in Sect.~4.

\section{Theoretical perspectives}

\subsection{Basic lensing notation}

We first outline the notation to be used below. As usual, lenses are
assumed to be  sheet-like perpendicular to the line-of-sight. Then,
the lens mapping from the lens to the source plane can be written as
\begin{equation} \label{eq:lens}
  \vec S=\vec L-\nabla\psi\;,
\end{equation}
and the local distortion of the lensed images is determined by the
Jacobian matrix
\begin{equation} \label{eq:metric}
  \mathcal{A}_{ij}=\frac{\partial S_i}{\partial L_j}\;,
\end{equation}
where $(S_1,S_2)$ and $(L_1,L_2)$ are dimensionless coordinates in
the source and lens planes,  respectively, and $\psi$ is the
effective lensing potential, which characterizes the lensing mapping
completely for any given lens model. The magnification at a given
position $\vec L$ is
\begin{equation} \label{eq:mag}
  \mu=\frac{1}{\det\mathcal{A}}\;.
\end{equation}
Giant arcs appear close to the critical curves when their sources
are close enough to  the corresponding caustics. Since giant arcs
are the most pronounced structures due to highly non-linear strong
lensing, they provide sensitive constraints on the lens models.

For simplicity, we adopt in the following subsections a
one-component lens model to  illustrate the predicted magnifications
of expected giant arcs for the SIS, NIS, and NFW profiles,
respectively. Although real lenses typically have multiple
components, the main conclusions on the magnification will remain
valid because they are due to generic lensing behaviour near
critical curves or caustics.

\subsection{Axial symmetry}

\begin{figure}
\includegraphics[width=\hsize]{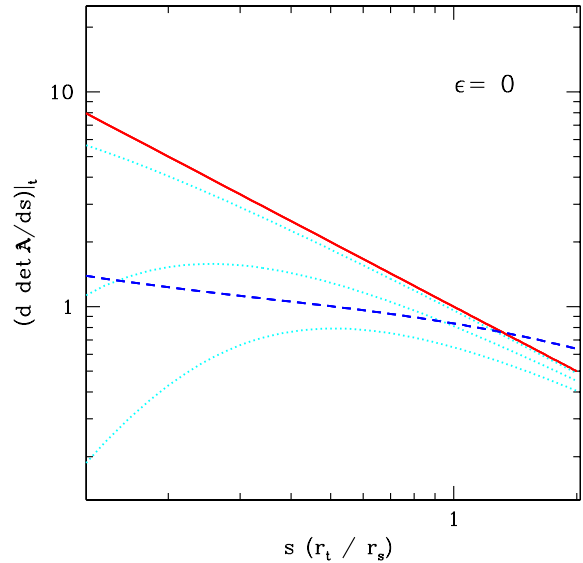}
\caption{The predicted gradient $(\d\det\mathcal{A}/\d s)|_t$ of the
Jacobian determinant at the position of the tangential critical
curve is shown as a function of the dimensionless radius $s$ with
the solid, dotted and dashed lines for the SIS, NIS, and NFW density
profiles, respectively. Dotted lines from top to bottom show the
results for the NIS model with $r_\mathrm{s}/r_\mathrm{c}=50, 10,
5$, respectively, compared to the scale radius of the NFW model (see
the text for details).} \label{sym_dadr}
\end{figure}

If the lens is axially symmetric, i.e.~the ellipticity $\epsilon=0$,
we can obtain the radial gradient $(\d\det\mathcal{A}/\d s)|_
\mathrm{t}$ of the lens mapping Eq.~(\ref{eq:metric}) at the radius
$r_\mathrm{t}$ where giant tangential arcs appear. There,
$\det\mathcal{A}=0$. We introduce
\begin{equation}
  s=r/r_\mathrm{s}
\end{equation}
throughout this section for easier comparison. We plot the gradient
$(\d \det\mathcal{A}/\d s)|_{\mathrm{t}}$ as a  function of $s$ in
Fig.~\ref{sym_dadr}, where the solid, dotted and dashed lines show
the results for SIS, NIS, and NFW profiles, respectively.

The radial gradient of the Jacobian determinant at the tangential
critical  curve is smaller for NIS profiles with different core
radii than for the SIS profile, since the NIS density profile is
always flatter than that of the SIS, thus the projected mass
distribution is flatter for the NIS. With increasing core radii in
NIS profiles, i.e.~decreasing $r_\mathrm{s}/r_\mathrm{c}$, the
gradient decreases due to the flattening of the density profiles. If
the giant arc is located at a radius $r_\mathrm{t}\ga r_\mathrm{s}$,
i.e., $s\ga 1$, the gradient is larger for a NFW than for the SIS
and NIS profiles because it is steepens towards $\propto r^{-3}$
outside the scale radius. If the giant arc appears in the inner
region, i.e.~$r_\mathrm{t}\la r_\mathrm{s}$, the gradient for a SIS
or a NIS lens with a small core ($r_\mathrm{s}/r_\mathrm{c}\ga10$)
is larger than that for a NFW lens, while the gradient for a NIS
profile with a relative large core
($r_\mathrm{s}/r_\mathrm{c}\la10$) is smaller than for a NFW profile
due to its flatter density profile.

It is interesting to study the corresponding magnifications of the
giant  arcs produced by these different mass profiles. The
determinant $\det\mathcal{A}$ of the lens mapping in
Eq.~(\ref{eq:metric}) at a distance $\delta r$ from the critical
curve can be written as
\begin{equation} \label{eq:dadr}
  {\rm det}\mathcal{A}={\rm det}\mathcal{A}|_\mathrm{t}+
  \left.\frac{\d{\,\rm det}\mathcal{A}}{\d r}\right|_\mathrm{t}\delta r=
  \left.\frac{\d{\,\rm det}\mathcal{A}}{\d r}\right|_\mathrm{t}\delta r\;,
\end{equation}
where the subscript $t$ again denotes the position of the tangential
critical curve where $\det\mathcal{A}=0$. If the width of the arc is
$W$, its total magnification $\bar\mu_\mathrm{t}$ can be expressed
by
\begin{eqnarray} \label{eq:mu_s}
  {\bar \mu_\mathrm{t}}&=&\frac{\int\d S_i}{\int\d S_s}=
  \frac{\int\d S_i}{\int\d S_i/\mu_i} \nonumber \\
  &=&\frac{\int_{r_\mathrm{t}-{\rm W/2}}^{r_\mathrm{t}+{\rm W/2}}\;r\d r}
  {\left.\d\det\mathcal{A}/\d r\right|_\mathrm{t}
  \int_{r_\mathrm{t}-{\rm W/2}}^{r_\mathrm{t}+{\rm W/2}}\;(r-r_\mathrm{t})r\d
  r} \nonumber \\
  &=&\frac{12s_\mathrm{t}}{\d\det\mathcal{A}/\d s|_\mathrm{t}}\left(
    \frac{r_\mathrm{s}}{W}
  \right)^2
  =\frac{12}{\d\det\mathcal{A}/\d s|_\mathrm{t}s_\mathrm{t}}\left(
    \frac{r_\mathrm{t}}{W}
  \right)^2\;.
\end{eqnarray}
If an observed giant arc appears at a radius $r_\mathrm{t} \sim
r_\mathrm{s} \sim15''$  from the cluster centre with a width of
$W\sim3''$, say, its estimated magnification can reach $\sim100$.

\begin{figure}
\includegraphics[width=\hsize]{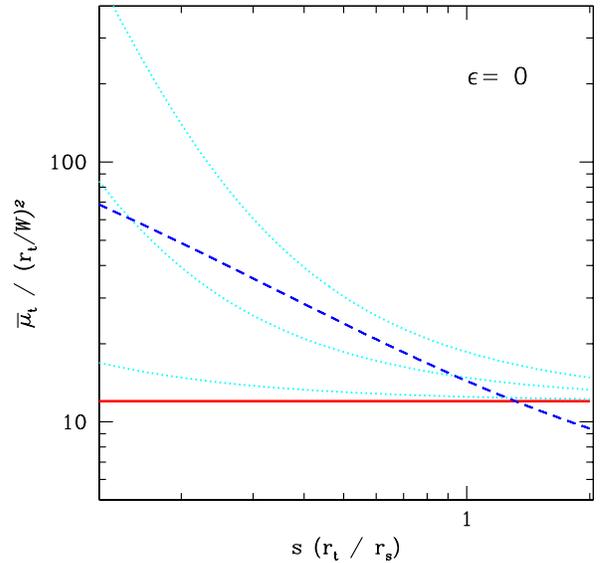}
\caption{The expected magnification for giant arcs,
$\mu_\mathrm{t}$, scaled by $(r_\mathrm{t}/W)^2$, as a function of
the dimensionless radius $s$. As in Fig.~\ref{sym_dadr}, the solid,
dotted and dashed lines represent the SIS, NIS, and NFW density
profiles, respectively. From bottom to top, the dashed lines show
the results for a NIS profile with $r_\mathrm{s}/r_\mathrm{c}=50,
10, 5$, respectively (see the text for more detail).}
\label{sym_mag}\end{figure}

The predicted total magnification $\bar \mu_\mathrm{t}$ scaled by
$(r_\mathrm{t}/W)^2$ for  giant arcs as a function of their distance
$s$ from the lens centre is plotted in Fig.~\ref{sym_mag} with the
same meaning of the line types as in Fig.~\ref{sym_dadr}. The figure
shows that the expected magnification of an arc is always lower for
a SIS than for a NIS model, given the observed positions of giant
arcs and their widths. This is because the SIS profile is always
steeper than the NIS profile. If the arc appears at $r_\mathrm{t}\ga
r_\mathrm{s}$, the predicted magnification for a NFW profile is the
lowest due to the steepest density distribution. If the arc appears
at $r_\mathrm{t}\la r_\mathrm{s}$, the expected magnification for
the SIS or the NIS models with a small core ($r_\mathrm{s}/r_
\mathrm{c}\ga10$) is lower than that for a NFW model. Only when the
core of the NIS profile is large ($r_\mathrm{s}/r_\mathrm{c}\la10$),
comparable to the scale radius of a NFW profile, the expected
magnification is larger than that for a NFW model since its density
profile will be significantly flatter in the inner region. The
conclusions can thus be well understood in terms of the flatness of
the density profile near the arc locations. See also the discussion
in Meneghetti et al (2005b).

\subsection{Asymmetry}

Since the gravitational potentials of real lensing clusters must be
asymmetric,  as was already pointed out by Grossman \& Narayan
(1988), Kovner (1989), Mellier et al (1993) and Miralda-Escud\'e
(1995), we now turn to asymmetric models, which play an important
role in cluster-lensing studies (Bartelmann \& Meneghetti 2004). As
in Bartelmann \& Meneghetti (2004), we define the surface-mass
distribution of the cluster in terms of the projected elliptical
radius
\begin{equation} \label{eq:re}
  r_\mathrm{e}=\left[
    (r\cos\theta)^2(1-e)+(r\sin\theta)^2/(1-e)
  \right]^{1/2}\;.
\end{equation}
Here, $\theta$ is the position angle counterclockwise from the $+y$
axis.  The ellipticity
\begin{equation} \label{eq:e}
  e=\frac{a-b}{a+b}
\end{equation}
of an ellipse with major and minor axes $a$ and $b$ refers to the
lensing potential rather than the projected density. Note that the
definition is chosen so as to conserve the total projected mass
within $r_\mathrm{e}$ independent of ellipticity, as pointed out by
Meneghetti et al (2005b).

Projected ellipticities of real clusters are not very large; they
are always smaller than their real ellipticities in three
dimensions. We restrict our theoretical results to the first order
in $\epsilon$ in this subsection. Then, the critical curve
approximates an ellipse. Without restriction, we assume that its
major and minor axes are oriented along the $x$ and $y$ coordinates,
respectively. Giant arcs preferentially appear near positions with
either $x\approx0$ or $y\approx0$ from sources in the neighbourhood
of cusp points in the caustics.

\begin{figure}
\includegraphics[width=\hsize]{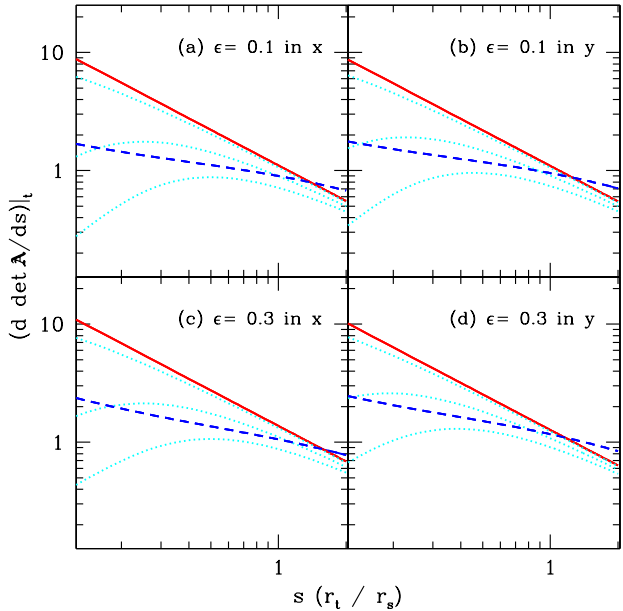}
\caption{Same as Fig.~\ref{sym_dadr}, but choosing different
ellipticities $\epsilon\ne0$. Giant arcs appearing near the $x$ and
$y$ axes are considered (see text for further detail).}
\label{asym_dadr}
\end{figure}

The gradient of the Jacobian determinant $(\d\det\mathcal{A}/\d
s)|_\mathrm{t}$ of the lens mapping Eq. (\ref{eq:metric})
perpendicular to the critical curve at $x=0$ and $y=0$, where giant
tangential arcs are expected to appear, is shown in
Fig.~\ref{asym_dadr} as a function of the dimensionless radius $s$
for different ellipticities $\epsilon$. The line types are chosen as
in Figs.~\ref{sym_dadr} and \ref{sym_mag}. For a given $\epsilon$,
the gradient exhibits the same behaviour as in Fig.~\ref{sym_dadr}
for axially-symmetric SIS, NIS, and NFW density profiles for the
same reasons.

Comparing with axially-symmetric lens models, the situation for
different  ellipticities $\epsilon$ is more complicated. Two factors
will affect the gradient of the Jacobian determinant
$(\d\det\mathcal{A}/\d s)|_\mathrm{t}$ and its influence on a giant
arc at a given position: (1) the total mass enclosed by a circle
traced by the arc, and (2) the density gradient. Ellipticity
enhances the tidal field and thus shifts critical curves towards
larger radii. Thereby, it increases the gradient
$(\d\det\mathcal{A}/\d s)|_\mathrm{t}$ because of the increasing
steepness of the projected density profile for a given mass model.

\begin{figure}
\includegraphics[width=\hsize]{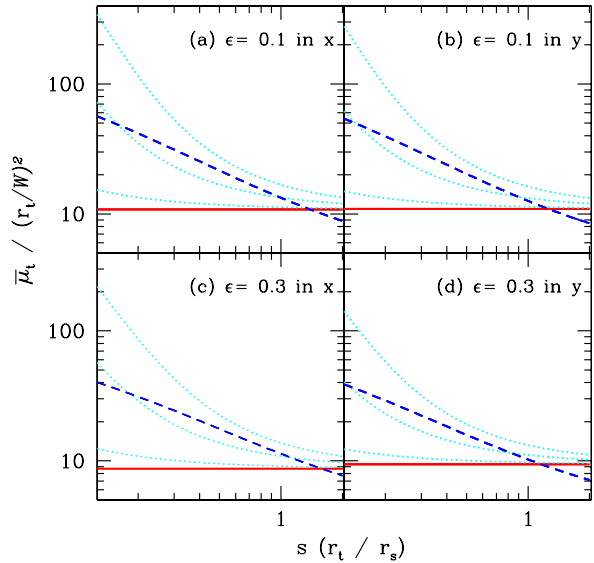}
\caption{Similar to Fig.~\ref{sym_mag}, but adopting different
ellipticities $\epsilon$ of the lensing potential. Results are shown
for giant arcs appearing near the $x$ and $y$ axes, respectively
(see the text for more detail).} \label{asym_mag}
\end{figure}

As in the previous subsection, we plot in Fig.~\ref{asym_mag} the
expected  magnification for arcs near the $x$ and $y$ axes as a
function of their distance from the cluster centre for asymmetric
lens models. The figure shows that the arc magnifications show the
same general trends as in Fig.~\ref{sym_mag}, for the same reasons.
Note that the magnifications of axially-symmetric and
axially-asymmetric models differ significantly even if the
ellipticity $\epsilon$ is small. The predicted magnifications of
arcs for individual mass models decrease for increasing $\epsilon$
because of similar reasons as discussed above, especially for arcs
near $x = 0$, i.e.~near the major axis.

In summary, the expected magnifications for giant arcs in lenses
with different  mass models are different even though the arc
positions can be reproduced by all of them. Asymmetry of the mass
distribution plays an important role for strong lensing. The
resulting magnifications can change substantially even if the
system's ellipticity is small. Although the absolute magnifications
for individual arcs or arclets cannot be obtained observationally,
lens models will be additionally constrained by the relative
magnifications of arc or arclet pairs from the same sources within a
lens system. We shall use the well-studied clusters A~370 and
MS~2137 as illustrations in the next section.

\section{Illustration: A~370 and MS~2137}

As discussed in the previous section, different mass models are
possible for individual galaxy clusters if they are only based on
the observed positions of giant arcs. Relative magnifications of arc
or arclet pairs from individual sources in a lens system can be used
to constrain the mass model further. We will investigate the arc
magnifications in NIS and NFW models, which can well reproduce the
arc configurations in A~370 and MS~2137, respectively. We choose
these two clusters for the following reasons:

\begin{itemize}

\item They are prototypical lensing clusters, with A~370 consisting of
two mass components and MS~2137 having one dominant component.

\item Both of them have large tangential and radial arcs, which mass models are required to fit.

\item Both of them have at least two arc or arclet pairs with
photometric data available in several bands, which allow us to use
relative magnifications (i.e.~flux ratios) of image systems from
individual sources to further constrain the mass models.

\item High-quality HST images are available for both of them, which can
be used for detailed strong-lensing studies.

\item Both clusters have been thoroughly studied earlier, which allows
detailed comparisons.

\item For A~370 in particular, we have high-quality images in the $B$,
$V$, $R$, $I$, $z$, $J$, $H$ and $K'$ bands taken by Subaru with a
median seeing of $\sim0''.5$ (Cowie et al. 2008), and in four
infrared channels observed by IRAC on-board the \emph{Spitzer} space
telescope in addition to its HST archive images. The deep optical
and NIR images permit the identification of individual images of
lensed galaxies in the A370 region studied in the present paper. The
IRAC images have $\gtrsim2''$ resolution and will be used to study
in detail the spectral energy distributions (SEDs) for the sources
combined with the optical and NIR data in a forthcoming paper.

\end{itemize}

We shall very briefly summarize earlier work on these two clusters
in the following subsections. Different mass models will be applied
to the two clusters in order to reproduce the main observed features
of arcs and arclets. Since there are many arclets observed in the
field of A~370 and extensive observational data are available, we
will study this cluster in more detail.

Our models improve the previous lens models for MS~2137 and A~370 in
the following way. Gavazzi et al (2003) studied the density profile
for MS~2137 in detail and Comerford et al (2006) successfully
applied the NFW density profile to this cluster. Both studies did
not explicitly consider the relative magnifications between image
pairs from individual sources as constrained by the photometric
observations, which we focus on in this paper. For A~370, we not
only employ more observed arclet pairs than Comerford et al (2006)
to constrain the mass model, but also compare their magnifications
with the observed photometry. Moreover, we study additional arclets
in more detail, including their estimated lens redshifts.

We use the same code as in Comerford et al. (2006) to fit mass
models to observations. Although only NFW profiles were fit to
observations there, isothermal density profiles are also provided by
this code.

Briefly, it proceeds as follows: Varying the density-profile
parameters, it strives (1) to make arc sources as small as possible,
(2) to make arcs and arclets produced from these sources match the
observed data as closely as possible, and (3) to avoid additional
images arising from the sources or make them faint enough to be
compatible with the observations.

These three requirements are quantified by three contributions to a
figur-of-merit function, which is jointly minimized to find the
best-fitting model parameters. This function is not precisely a
$\chi^2$ function, as it cannot be normalised to the degrees of
freedom because the intrinsic source size is unknown. See Comerford
et al. (2006) for more detail. Relative magnifications, i.e.~the
flux ratios, between images in image systems of individual sources,
are not included in the figure-of-merit function when fitting mass
models to MS~2137 and A~370 in the following subsections. They are
used in the further discussions on how well the best-fitting models
can match the photometric observations.

\subsection{MS~2137}

MS~2137 is a rich cD cluster at $z=0.313$. Early studies by Mellier,
Fort \& Kneib (1993) and Miralda-Escud\'e (1995) found that
elliptical isothermal models can reproduce the observed arc system
remarkably well. Miralda-Escud\'e (2002) suggested that its mass
distribution should be highly elliptic. Comerford et al (2006)
showed that the cluster can also be modelled with NFW components.
Based on archival data from the HST and the ESO-VLT, Gavazzi et al
(2003) and Gavazzi (2005) investigated its mass profile in detail
and found that it can be fit by the isothermal and the NFW density
profiles alike. Based on a possible fifth image in the very cluster
centre and on stellar kinematical data, they concluded that the
isothermal model may be preferred.

\begin{figure}
\includegraphics[width=\hsize]{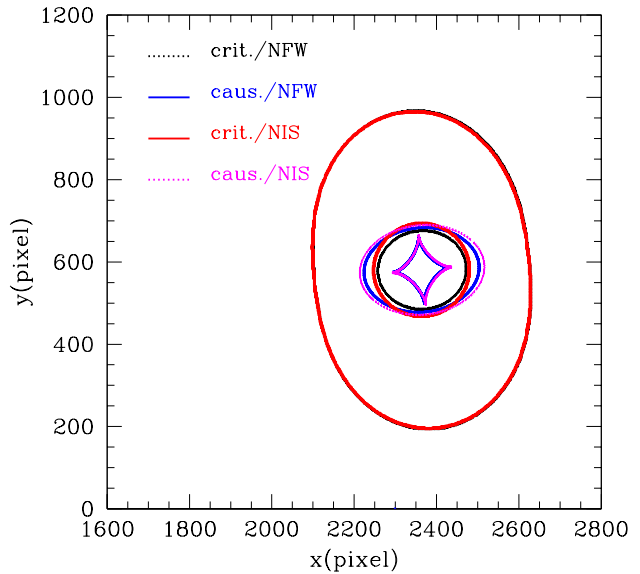}
\caption{Critical curves and caustics for the two best-fitting mass
models of MS~2137.} \label{ms2137 crit caus}
\end{figure}

We fit both the NIS and the NFW mass density profiles to MS~2137,
attempting to reproduce the observed five arcs including the radial
arc. We adopt a single lens component at the position of the cD
galaxy, similar to Comerford et al. (2006). The predicted critical
curves and caustics of the different best-fitting models are shown
in Fig.~\ref{ms2137 crit caus}. They are almost identical for the
two types of model, which confirms the results of earlier studies,
since a one-component lens system cannot be tightly constrained with
arc systems at identical or similar redshifts.

\begin{figure}
\includegraphics[width=\hsize]{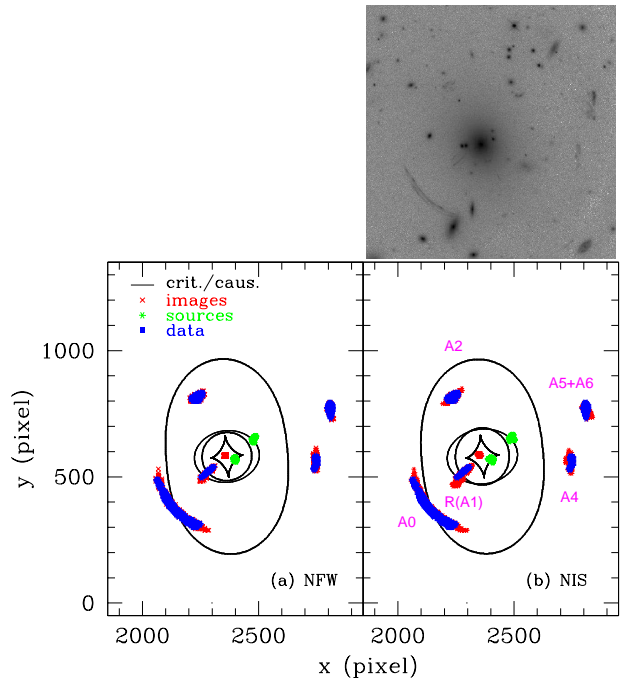}
\caption{The HST image (upper panel) and the reproduction of the
observed data (lower panel) by the two best-fitting mass models for
MS~2137, with (A0, A2, A4) and (A1, A5+A6) being identified as
counter-images of two different sources. Observed arcs, reproduced
images and the reconstructed sources are marked by $\diamond$,
$\times$, and $\ast$, respectively. The critical curves and caustics
of the models are marked in black.} \label{ms2137 arc}
\end{figure}

The reproduction of the arc systems by the best-fitting mass models
are shown in Fig.~\ref{ms2137 arc} adopting the notation of Gavazzi
et al (2003), where (A0, A2, A4) and (A1, A5+A6) are identified as
counter-images of two different sources. The best-fitting model
parameters are given in Tab.~\ref{tab_fitting}. The figures-of-merit
obtained with the complete set of input data points read off the
observational data are also listed in the table. It appears that the
NFW profile is preferred based on the figures-of-merit of different
best-fitting mass models. Note that the figure-of-merit in this
paper adds the three contributions of sources, the observed and the
predicted images. Although it does not have the usual absolute
statistical meaning of a properly normalised $\chi^2$ function, it
can still serve to evaluate different models based on identical
input data (see Comerford et al (2006) for more detail).

Since we use the same multiple-image systems as Comerford et al
(2006) to constrain our models, the model parameters obtained for
the NFW profile are similar except for a very small difference in
$\kappa_\mathrm{s}$, where we find $0.66$ rather than the $0.67$ of
Comerford et al (2006), which is perfectly within the expected
uncertainty.

The parameter $\kappa_\mathrm{s}$ is the amplitude of the
convergence profile $\kappa(x)$, defined by
\begin{equation}
  \kappa(x)=\frac{2\kappa_\mathrm{s}}{x^2-1}\left[
    1-\frac{2}{\sqrt{x^2-1}}\arctan\sqrt{\frac{x-1}{x+1}}
  \right]
\label{kappa_NFW}
\end{equation}
for the NFW profile with $x=r/r_\mathrm{s}$ and by
\begin{equation}
  \kappa(x)=\frac{\kappa_\mathrm{s}}{2}
  \frac{x^2+2x_\mathrm{c}^2}{(x^2+x_\mathrm{c}^2)^{3/2}}
\label{kappa_NIS}
\end{equation}
for the NIS profile, where $x$ and $x_\mathrm{c}$ are the radius $r$
and the core radius $r_\mathrm{c}$ in units of the Einstein radius.
Both the NFW and the NIS models predict a fifth-image in the central
cluster region which was confirmed by Gavazzi et al (2003) after
subtracting the light produced by the cD galaxy.

\begin{table}
\caption{Parameters of the best-fitting models for MS~2137 and
A~370, with Column (1): name of the cluster; (2): label of lenses
where applicable; (3): convergence amplitude $\kappa \mathrm{s}$;
(4) scale radius for NFW, core radius for NIS models; (5):
ellipticity; (6): position angle; (7): averaged figure-of-merit for
the total number of data points; (8): adopted mass model.}
\begin{tabular}{llrrrrrl}
\hline\hline cluster & lens & $\kappa_\mathrm{s}$ & $r_\mathrm{c/s}$
               & $\epsilon$ & $\theta$ & $\chi^2$ & model \\
& & & ($\mathrm{kpc}/h$) & & (deg) &  &\\
\hline
MS~2137 & -- & $0.66$ &  $64$ & $0.11$ &  $95.7$ & 0.19 & NFW \\
        & -- & $3.12$ &  $17$ & $0.11$ &  $95.7$ & 0.82 & NIS \\
\hline
A~370   & G1 & $0.16$ & $254$ & $0.28$ &  $78.0$ & 0.16 & NFW \\
        & G2 & $0.17$ & $212$ & $0.07$ & $-13.4$ & &NFW \\
\hline
        & G1 & $50.0$ &   $1$ & $0.20$ &  $75.0$ & 2.5& NIS \\
        & G2 & $30.0$ &   $1$ & $0.05$ & $-10.0$ & & NIS \\
\hline\hline
\end{tabular}
\label{tab_fitting}
\end{table}

\begin{figure}
\includegraphics[width=\hsize]{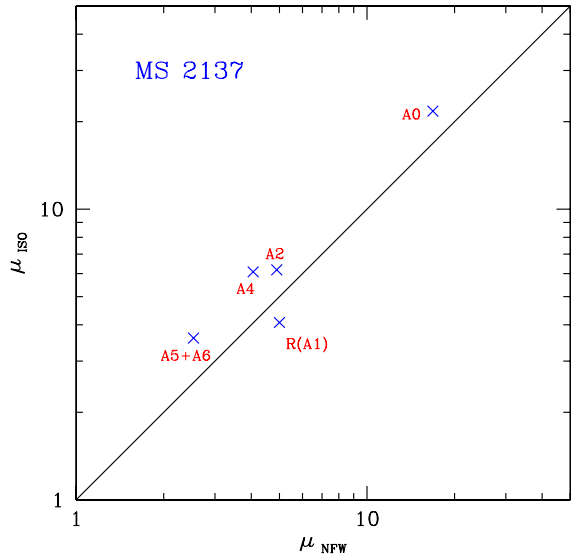}
\caption{Predicted magnifications of arcs and arclets in the fields
of MS~2137 based on the NIS model, compared to those based on the
NFW model, with (A0, A2, A4) and (A1, A5+A6) being identified as
image systems of two different sources. The solid line illustrates
where they would agree.} \label{compare 2137}
\end{figure}

The predicted absolute magnifications for the individual arcs are
shown in Fig.~\ref{compare 2137} for the two different best-fitting
mass models for MS~2137, with the solid diagonal indicating where
the two magnifications for individual arcs would agree. As the
figure shows, the arc magnifications predicted by the two models
agree well. This implies that further constraints on the density
profile for MS~2137 based on relative image magnifications of arcs
from individual sources cannot be expected because MS~2137 is a
relatively simple lens.

\begin{figure}
\includegraphics[width=\hsize]{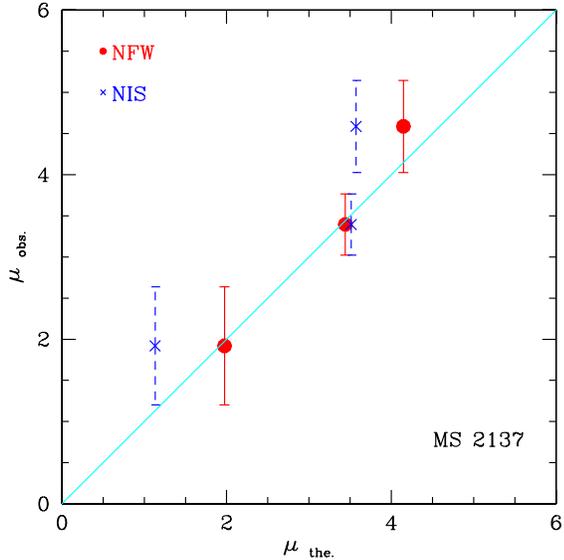}
\caption{Comparison of the predicted relative magnifications of arcs
and arclets pairs in the fields of MS~2137 with the observations.
Points from left to right denote R(A1)/A5+A6, A0/A2 and A0/A4 with
the photometric errors attached. The solid diagonal indicates where
they would agree.} \label{obs_2137}
\end{figure}

Observationally, we can obtain the relative magnifications for three
arc or arclet pairs, i.e.,  A0/A2 and A0/A4 for the giant arc and
A1/(A5+A6) for the radial arc, resulting from two sources based on
the photometric data in seven bands ($U$, $B$, $V$, $R$, $I$, $J$
and $K$; Gavazzi et al 2003). The averaged results obtained in seven
bands are shown in Fig.~\ref{obs_2137} with their photometric
errors. We find that the model predictions by both the NIS and the
NFW model are roughly in agreement with the averaged relative
magnifications for these three pairs within the observational
errors, although the NFW model again appears slightly preferred. We
conclude that both the NIS and the NFW mass profiles work well for
MS~2137, even if the magnification, which is the second order of the
derivative of the potential, is considered.

Moreover, with the exception of the radial arc, the predicted
magnifications are somewhat higher for the NIS than for the NFW
model, which can be understood as follows. According to the
parameters of the best-fitting models for MS~2137 given in
Tab.~\ref{tab_fitting}, $r_\mathrm{s}/r_\mathrm{c}\sim4$, showing
that the core radius of the NIS model is comparable to the scale
radius of the NFW profile. This yields higher predicted
magnifications for the arcs reproduced by the NIS compared to the
NFW model (cf. Fig.~\ref{asym_mag}). It should be pointed out that
the predicted magnification of the radial arc is calculated
according to Eq.~(\ref{eq:mag}) based on its observed image
location, which straddles the corresponding critical curve. The
resulting magnification mainly depends on its observed size, which
is difficult to measure accurately since it is located in the very
central region of the lens.

\subsection{A~370}

Abell~370 ($z=0.375$) was the first cluster in which gravitational
arcs were observed (Soucail et al 1987). This two-component,
strongly lensing system provides an interesting case for us to study
the magnifications of individual observed arcs and arclets in
detail, and to compare with previous work.

This cluster has frequently been studied. For instance, Kneib et al
(1993) reconstructed the mass distribution of A~370 based on the
observed giant arc (A0) and the arc pair (B2,B3) to conclude that
the dark matter appears to follow the light distribution for this
cluster. Bartelmann (1996) showed that, as in MS~2137, the radial
arc can be reproduced by an NFW model, which was recently confirmed
with a detailed model by Comerford et al (2006). Using a deep HST
image, B\'{e}zecourt et al (1999) constructed a mass model for A~370
based on a truncated pseudo-isothermal distribution and found that
the observed lensed images can be well reproduced. Using A~370 as a
natural telescope, Hu et al (2002) found a galaxy at $z\approx6.56$
and Chary, Stern \& Eisenhardt (2005) argued that its star formation
rate could be $\ga140\,M_\odot\,\mathrm{yr}^{-1}$, thus providing
useful clues to the galaxy-formation models at high redshift.

\begin{figure}
\includegraphics[width=\hsize]{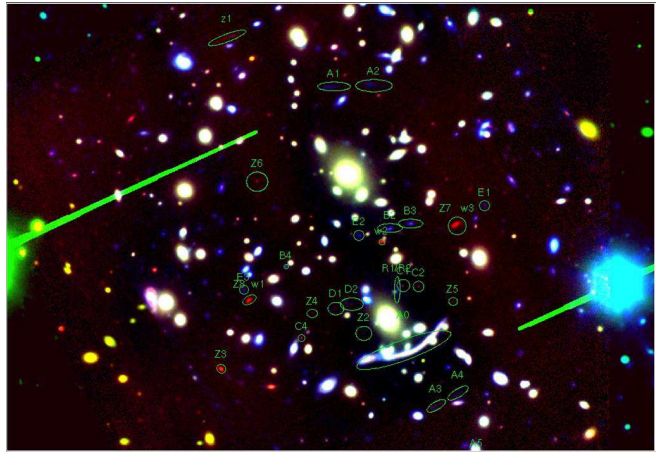}
\caption{Composite image of A~370 taken in the $B$, $R$ and $K'$
bands with the arcs and arclets marked.} \label{a370color}
\end{figure}

We adopt the notation of B\'{e}zecourt et al (1999) and Kneib et al
(1993) for the different arcs and arclets in the field of A~370 to
facilitate comparisons with earlier studies. The additional arclet
candidates we identify are jointly labeled by ``Z'' below. For
clarity, we only show one colour image of A~370 combined from the
$B$, $R$ and $K'$ bands rather than separate images in individual
bands, with the individual arcs and arclets identified in
Fig.~\ref{a370color}. Note that A5 appears near the bottom since it
appeared near the boundary of the $K'$-band image, but it can be
very clearly seen in the $B$ band.

Since there are many arcs and arclets in the field of A~370, we list
below the image systems produced by individual sources for clarity,
except the well defined giant arc A0, the radial arc R and the arc
systems B, C, D and E (B\'{e}zecourt et al 1999). In the labeled
``A" sequence, (A1, A2) and (A3, A4) are arc pairs produced by two
different sources. A5 is a single arc. For the newly found arcs
labeled by ``Z", we suggest in the next subsection that Z1, Z3, and
Z6 be arcs produced by three different sources and (Z2, Z4, Z5)
result from a single source. Z7 and Z8 (also labeled as W3 and W1
respectively in Fig. \ref{a370color}) with W2 are identified as
counter-images of another source (see next subsection).

As for MS~2137, we first construct both a NIS and a NFW mass model
for A~370 reproducing the giant arc A0, the radial arc R, the
arclets (A1, A2), (B2, B3), and A5. An extra arclet pair labeled
(A3, A4) is also included, which is easily seen in
Fig.~\ref{a370color} especially in the $B$-band. In total, we use
nine relatively bright arcs and arclets produced by six sources to
constrain the mass model for A~370. The lens components are fixed at
the centres-of-light of the two bright cD galaxies.

\begin{figure}
\includegraphics[width=\hsize]{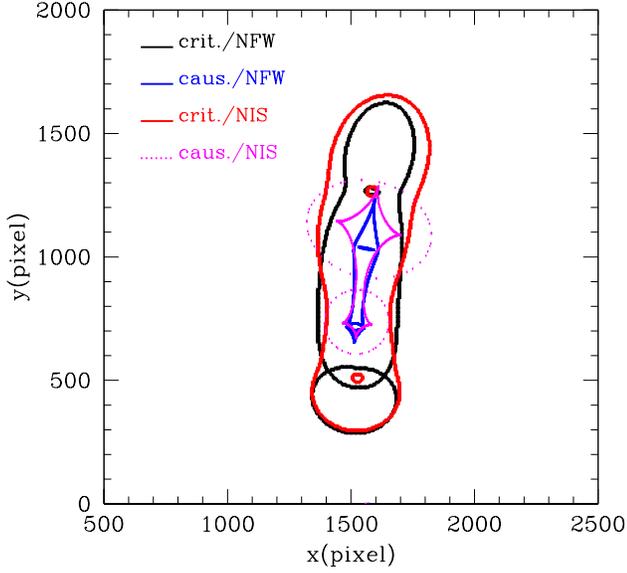}
\caption{The critical curves and caustics for the best-fitting mass
models for A~370.} \label{a370 crit caus}
\end{figure}

\begin{figure}
\includegraphics[width=\hsize]{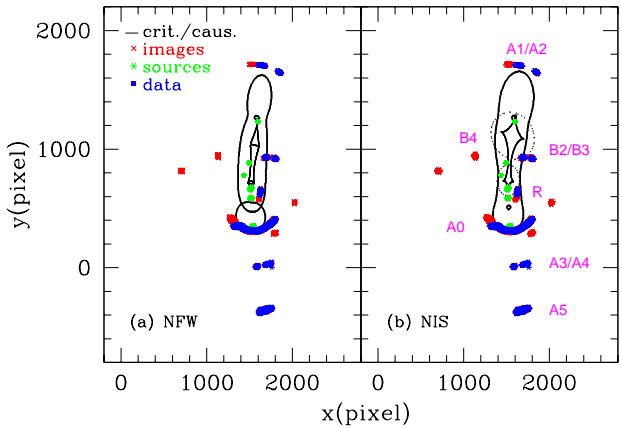}
\caption{The best-fitting results for A~370, using the same notation
as in Fig.~\ref{ms2137 arc}.} \label{a370 9arc}
\end{figure}

The best-fitting model parameters are listed in
Tab.~\ref{tab_fitting} together with the averaged figures-of-merit
of the complete set of data points. The corresponding critical
curves, caustics and the reproduced images are shown in
Figs.~\ref{a370 crit caus} and \ref{a370 9arc}. Our results for the
NFW model are very close to those found by Comerford et al (2006)
where the six arcs and arclets A0, R, (A1, A2) and (B2, B3) were
taken into account. This suggests that (A1, A2) and (B2, B3) play
important roles in the mass configuration for A370 (Kneib et al
1993). We point out that the high value of $\kappa_\mathrm{s}$ and
the low value of $r_\mathrm{c}$ in the best-fitting NIS model imply
that the arc in A~370 can be very well reproduced by nearly-singular
isothermal lens components.

Since the redshifts of individual sources can also be considered as
free parameters during the model fitting, we can obtain
gravitational-lensing estimates for individual source redshifts,
using the measured redshift of the giant arc A0 as a reference for
the relative lensing efficiency at different redshifts. This will be
discussed further below in this subsection.

Note that the critical curves and caustics of the best-fitting
models are very different for the two mass models, although both of
them reproduce the observed positions of individual arcs and arclets
very well (cf.~Fig.~\ref{a370 9arc}). Since the selected arcs and
arclets can well constrain the mass configuration for A~370, this
implies once again that mass profiles in clusters cannot be clearly
distinguished based exclusively on the image positions observed in
strong lensing.

Although both best-fitting models succeed in reproducing the main
features of the observed arcs and arclets, there are four extra
images predicted that were not yet found observationally. The image
near the top in Fig.~\ref{a370 9arc} is the counter-image predicted
for B2 and B3, which matches the observed counter-image labeled B4
in the figures of the next subsection. The other three are the
predicted counter-images of the radial arc. These deviations arise
because (1) the lens model constructed for A~370 only has two
components centred on the cD galaxies without taking perturbations
by additional lens components centred on less massive cluster
galaxies into account; (2) the three additional images are too faint
to be observed which is consistent with the predictions that their
magnifications are small.

\begin{figure}
\includegraphics[width=\hsize]{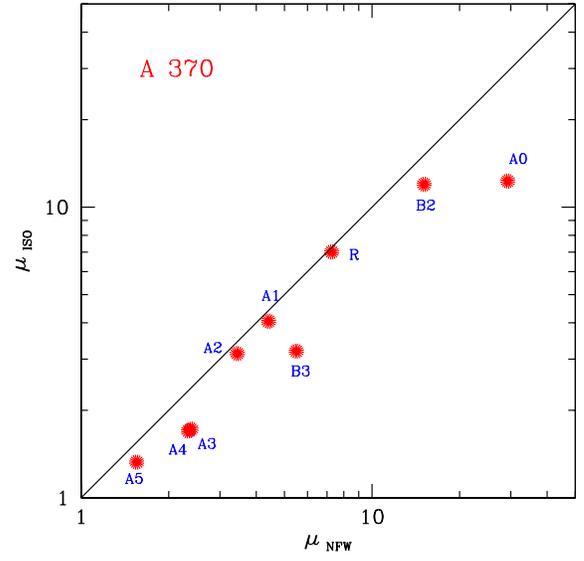}
\caption{Predicted magnifications of arcs and arclets in the field
of A~370 based on the NIS model, compared to those based on the NFW
model. The solid diagonal indicates where they would agree.}
\label{compare_370}
\end{figure}

\begin{figure}
\includegraphics[width=\hsize]{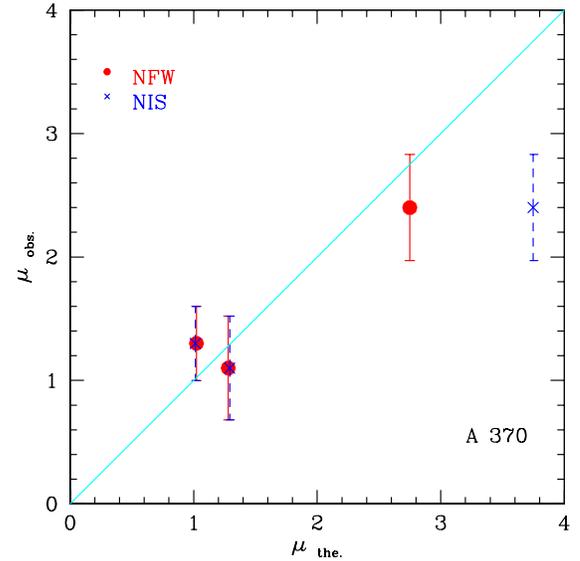}
\caption{Comparison of the predicted relative magnifications of arcs
and arclets pairs in the field of A~370 with the observations.
Points from left to right denote A3/A4 , A1/A2 and B2/B3 with the
observational errors added, respectively. The solid diagonal
indicates where they would agree.} \label{obs 370}
\end{figure}

The predicted magnifications of the arcs and arclets selected to
constrain the mass model for A~370 are calculated for both the NIS
and the NFW density profiles and displayed in Fig.~\ref{compare_370}
for comparison. The expected magnifications of these nine arcs and
arclets are always larger for the NFW than for the NIS profile, with
the magnification of the giant arc being larger by a factor of
$\sim2.5$. This is easily understood based on the theoretical
results shown in Fig.~\ref{asym_mag}. Since the ratio
$r_\mathrm{s}/r_\mathrm{c}$ is very large for the best-fitting
models of A~370, the magnifications by the NFW profile must exceed
those by the NIS profile.

Since there are many arc and arclet pairs in A370 and high-quality
images are available in eight optical and NIR bands, we can compare
the expectations for the relative magnifications of the confirmed
arclet pairs (B2, B3), (A1, A2) and (A3, A4) with the observations
in individual bands, to determine which density profile may be
preferred. Photometric data reduction is done using SExtractor
(Bertin \& Arnouts 1996) for optical and NIR. Here, we only show the
results, while the detailed photometry will be described in a
forthcoming paper studying the SEDs of the lensed galaxies in A~370.

In Fig.~\ref{obs 370}, we plot the relative magnifications (i.e.,
flux ratios) of individual arclet pairs averaged over eight
independent bands against the corresponding predicted values. Unlike
in MS~2137, we find that the \emph{relative} magnifications expected
from the NFW profile can match the observations much better than
those based on the NIS profile. The pair (B2, B3) plays an important
role here. Mellier et al (1991) had already pointed out that (B2,
B3) is very sensitive to the lens model. We confirm that the
best-fitting NFW profile achieves a lower figure-of-merit than the
NIS profile. According to the above analysis and current numerical
simulations, we thus suggest that the NFW profile for A~370 matches
the strong-lensing observations better than the NIS profile, and
adopt it in the remainder of this paper.

As discussed above, the redshifts of individual sources can be seen
as input parameters in the model fitting because lensing by a fixed
mass distribution depends on a distance factor (see Comerford et al
2006) for sources at different redshifts. This allows redshift
estimates for individual sources required by the best-fitting
models. We proceed as follows. We choose the giant arc A0 as a
reference, which has the spectroscopic redshift $z_{\rm A0}=0.724$
(Soucail et al 1988). By varying the distance factors of the other
eight arcs and arclets and optimising the model, we estimate their
redshifts. The error bars are estimated by the redshift interval
within which the figure-of-merit increases by a factor of two from
the minimum.

We find that the source redshifts obtained in this way for the
best-fitting NFW and NIS models are similar. They are shown in the
upper part of Table~\ref{tab_z}. Redshifts from spectroscopy,
photometry and lensing estimates for the same sources obtained in
earlier studies are listed in Column 4 where available. The table
shows that the redshifts found here are consistent with the earlier
results except for the arc pair (A1, A2), which is suggested to have
a lensing redshift $z\approx1.4\pm0.2$ (B\'{e}zecourt et al 1999).
According to our photometric data in 12 bands, we find that A1 and
A2 are very blue (cf.~Fig.~\ref{a370color}) and are not detected by
IRAC. In qualitative agreement with our lensing-based estimate, this
suggests (A1, A2) are at lower redshift.

\begin{table}
\caption{Redshift estimates $z_\mathrm{l}$ (column 3) for arcs in
A~370 (column 1) obtained from distance factors (DF, column 2) in
the best-fitting lens model, compared to spectroscopic, photometric
or lensing-estimated redshifts, $z_\mathrm{s}$, $z_\mathrm{p}$ or
$z_{\rm l}$ (column 4), where available from the literature (column
5).}
\begin{tabular}{llllc}
\hline arc & DF & $z_\mathrm{l}$ & $z_\mathrm{s}$, $z_\mathrm{p}$ or
$z_{\rm l}$ &
ref. \\
\hline \hline
A0     & 1.00 & --   & 0.724 (s) & [1] \\
R      & 1.46$\pm$0.10 & 1.31$\pm$0.12 & 1.312 (p) & [2] \\
       &      &                & 1.7$\pm$0.2 (l) & [3]\\
B2,3   & 1.10$\pm$0.09 & 0.80$\pm$0.09 & 0.806 (s) & [3] \\
A1,2   & 1.15$\pm$0.11 & 0.85$\pm$0.11 & 1.4$\pm$0.2   (l) & [3] \\
       &      &      & $>$ 1.2  (p) & [4]\\
A3,4   & 1.30$\pm$0.12 & 1.02$\pm$0.11 & $>$ 0.2  (p) & [4] \\
A5     & 1.46$\pm$0.06 & 1.30$\pm$0.12 & 1.306  (s) & [5] \\
\hline \hline
B4     & 1.11 & 0.80 & --        & --  \\
C1,2,4   & 1.05$\pm$0.13 & 0.76$\pm$0.12 & 0.75$\pm$0.1  (l) & [3]\\
       &      &      & 0.60 - 0.92 (p)& [4]\\
D1,2   & 1.10$\pm$0.11 & 0.80$\pm$0.11 & 0.85$\pm$0.1  (l) & [3] \\
       &      &      & $<$ 0.4  (p) & [4]\\
E1,2,3 & 1.30$\pm$0.11 & 1.02$\pm$0.18 & 1.3$\pm$0.1   (l) & [3] \\
\hline \hline
Z1     & 1.70$\pm$0.05 & 2.28$\pm$0.29  & -- & -- \\
Z2     & 0.95$\pm$0.14 & 0.69$\pm$0.10  & -- & -- \\
Z3     & 1.60$\pm$0.06 & 1.73$\pm$0.21  & -- & -- \\
Z4     & 0.95$\pm$0.15 & 0.69$\pm$0.11  & -- & -- \\
Z5     & 0.95$\pm$0.14 & 0.69$\pm$0.10  & -- & -- \\
Z6     & 1.00$\pm$0.11 & 0.72$\pm$0.09 & -- & -- \\
Z7     & 1.34$\pm$0.08 & 1.08$\pm$0.11 & -- & -- \\
Z8     & 1.34$\pm$0.08 & 1.08$\pm$0.11  & -- & -- \\
W2     & 1.34$\pm$0.08 & 1.08$\pm$0.11  & -- & --\\
 \hline
\end{tabular}

{DF: distance factor (Comerford et al 2006); $z_{\rm l}$: lens
  redshift; $z_{\rm s}$: spectroscopic redshift; $z_{\rm p}$:
  photometric redshift; (s): spectroscopy; (p): photometry; (l): lens. }

{Refs: [1] Soucail et al (1988); [2] Smail et al (1996); [3]
B\'{e}zecourt et al (1999); [4] Kneib et al (1994); [5] Mellier et
al (1991)} \label{tab_z}
\end{table}

\subsection{Further study of the arclets in A~370}

The high-quality and deep images in eight bands from $B$ to $K'$
taken by Subaru enable us to apply the best-fitting NFW lens model
obtained in the preceding subsection to some arclets that were
already found and labeled by Kneib et al (1993) and B\'{e}zecourt et
al (1999), and to some new arclet candidates (labelled ``Z") in
order to test whether the best-fitting model is consistent with them
(see Fig.~\ref{a370color}).

\begin{figure}
\includegraphics[width=\hsize]{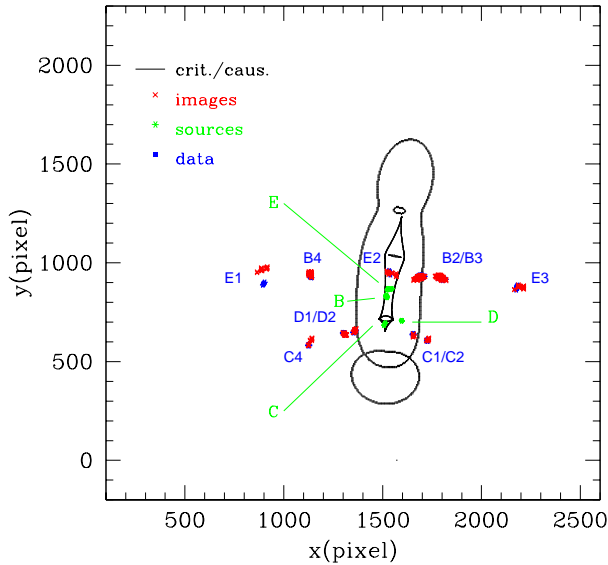}
\caption{The best-fitting results for the arclet pairs B, C, D, E in
A~370.} \label{a370 bcez}
\end{figure}

We thus apply the best-fitting NFW lens model obtained in the
previous subsection to the arclet pair candidates labeled B, C, D
and E by first optimising the model under the assumption of minimal
source sizes, only using their redshifts as free parameters. The
results of this fit for individual arclet pairs are shown in
Fig.~\ref{a370 bcez}, and their estimated redshifts are listed in
the middle part of Table~\ref{tab_z}.

Figure~\ref{a370 bcez} shows that the best-fitting NFW model can
fairly reproduce the observed images of the arclet pairs B, C, and
D. Note that the additional image produced by (B2, B3) in
Fig.~\ref{a370 9arc} matches well their possible counter image B4.
Their estimated redshifts are consistent with previous photometric
measurements. Small differences of the image positions between the
model predictions and the observations for E3 in the right panel of
the figure may arise because we ignore perturbations due to lensing
by less luminous galaxies within A~370. Such differences in the
image positions measured and reproduced by the model also occur for
C4. For the pair (D1, D2), the model predicts an extra counter-image
very close to D1 that is not clearly observed, possibly because the
image was yet too faint to be seen.

\begin{figure}
\includegraphics[width=\hsize]{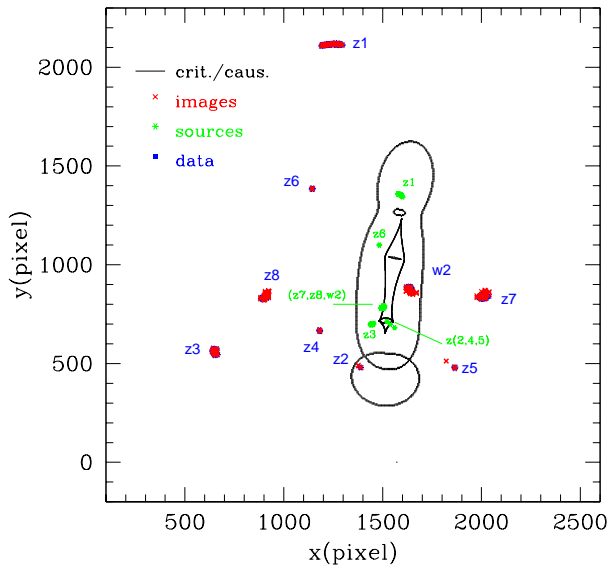}
\caption{The best-fitting results for the new arclet candidates in
A~370.} \label{a370_more_arcs}
\end{figure}

In addition to the arclet candidates labeled by Kneib et al (1993)
and B\'{e}zecourt et al (1999), we identify new arclet candidates in
the field of A~370 labeled ``Z1'' to ``Z8'' and shown in
Fig.~\ref{a370color}, with the measured ellipticities of individual
objects larger than 0.6 in  $R$ band by SExtractor (Bertin \&
Arnouts 1996) and checking by eye in the other bands. Similar to B,
C, D and E, the best-fitting NFW model is also applied to them,
taking their redshifts as free parameters. Best-fitting results for
them are shown in Fig.~\ref{a370_more_arcs} with the expected
lensing redshifts for individual new arclets being listed in the
final part of Table~\ref{tab_z}. A detailed study of their SEDs will
be presented in a separate paper.

Based on these results for the new arclets, we suggest that Z6 may
be a lensed galaxy at low redshift. Z2, Z4 and Z5 could be
counter-images because their recovered sources are close to each
other in the sky, and they have similar lensing redshifts. Although
the predicted separations of the source positions between Z4 and Z2,
and between Z4 and Z5 are possibly too large, this may be due to
model uncertainties. Moreover, two candidates of distant objects, Z1
and Z3, are found with lensing redshifts of $\sim2.3$ and
$\sim1.73$, respectively, which need to be spectroscopically or
photometrically confirmed.

Finally, the new candidates Z7 and Z8 are suggested to be
counter-images of a source at redshift near 1.3. Based on the
best-fitting NFW model, we also found their predicted counter image
close to B2, labeled W2 in Figs.~\ref{a370_more_arcs} and
\ref{a370color}. Photometry confirms that it has the same colour as
Z7 and Z8. We note that this newly found arclet pair had already
been detected by Bernard Fort and collaborators several years ago,
but remained unpublished (B.~Fort, private communication). Like (B2,
B3), these counter images are also very sensitive to the mass model
of A~370. Since they are extremely red, the study of their SEDs in
the next paper suggests that they arise from an active star-forming
galaxy heavily enshrouded by dust.

\section{Conclusions}

Since it is virtually impossible to distinguish mass models with
different density profiles for galaxy clusters based exclusively on
the observed arc and arclet positions, which depend on the first
derivative of the effective lensing potential, we have studied in
detail the image magnifications expected from lens models assembled
from components with isothermal and NFW density profiles. We
considered both axially-symmetric and asymmetric models and conclude
that the magnifications for different mass models can be
substantially different. The ellipticity of a mass model must be
taken into account because it plays an indispensable role in
strong-lensing studies.

Given a lens model reproducing the position of a giant arc, its
predicted magnification is usually smaller for singular than for
non-singular isothermal profiles. Isothermal models with small cores
predict magnifications smaller than models with NFW mass profiles.
Isothermal mass models with large cores, whose density gradient in
the inner region is nearly flat, predict magnifications increasing
with the core radius and exceeding those for NFW density profiles.
These effects are more pronounced for increasingly asymmetric
models.

We illustrate our results with the two well-studied clusters MS~2137
and A~370, which we model using the code described by Comerford et
al (2006). Comparing with previous studies, the main improvements in
this paper are based on (1) the photometry for obtaining relative
magnifications of individual counter-image pairs to constrain the
lens models; and (2) further studies of additional arclets in A~370
including lens redshifts of newly found arclets.

The observational data were taken with Subaru, HST and Spitzer for
A~370 and taken from Gavazzi et al (2003) for MS~2137. We find that
both isothermal and NFW models can reproduce the observed arcs and
arclet positions in MS~2137 and A~370, which is consistent with the
results of Comerford et al (2006), Gavazzi (2005), Gavazzi et al
(2003), Miralda-Escud\'e (2002, 1995), B\'{e}zecourt et al (1999),
Mellier, Fort \& Kneib (1993), and Kneib et al (1993).

Relative magnifications of individual arc and arclet pairs are
adopted to  adjust the mass model by comparing with observations.
Since MS~2137 is a relatively simple, single-component system, it is
hard to distinguish isothermal from NFW mass models. On the other
hand, A~370 has two major lens components with several arc and
arclet pairs, which allow us to adopt the relative magnifications of
the confirmed arclet pairs to conclude that the NFW profile is more
realistic than the isothermal profile for A~370.

The best-fitting NFW mass model for A~370 is applied to several
arclet candidates found by Kneib et al (1993) and B\'{e}zecourt et
al (1999) that were not used for constraining the mass model. We
find that this best-fitting model can well reproduce the observed
arclets.

The high-quality images of A~370 in twelve bands from $B$ to
infrared will allow us to recover SEDs for individual sources in the
field of A~370 in a forthcoming paper. We also find eight new arclet
candidates, apply our best-fitting NFW lens model to them and show
that we can reproduce their observed features fairly well. We obtain
redshift estimates from lensing for individual arcs and arclets in
A~370 that are consistent with spectroscopic and photometric results
where available from earlier studies. Other redshift predictions
from lensing need to be confirmed by future observations. Finally,
we found a new arclet pair labeled Z7, Z8 and W2 and suggest that
they arise from an active star-forming galaxy heavily enshrouded by
dust.

\acknowledgements

We thank Raphael Gavazzi for very helpful discussions. CS and BZ
acknowledge the financial support of the \emph{Deutsche
Forschungsgemeinschaft} for a visit to Heidelberg University and are
grateful to the Institute for Theoretical Astrophysics for its
hospitality during the visit. This project is partly supported by
the Chinese National Science Foundation No. 10333020, 10528307,
10778725, 973 Program No. 2007CB815402, Shanghai Science Foundations
and Shanghai Municipal Education Commission No.05DZ09. JMC
acknowledges support by a National Science Foundation Graduate
Research Fellowship.


\clearpage


\begin{thebibliography}{}
\bibitem{} Bartelmann, M., Meneghetti, M., 2004, A\&A, 418, 413

\bibitem{} Bartelmann, M., Meneghetti, M., Perrotta, F., Baccigalupi, C.,
 Moscardini, L., 2003, A\&A, 409, 449

\bibitem{} Bartelmann, M., Huss, A., Colberg, J. M., Jenkins, A., Pearce, F.
R., 1998, A\&A, 330, 1

\bibitem{} Bartelmann, M., 1996, A\&A, 313, 697

\bibitem{} Bertin, E., Arnouts, S., 1996, A\&A,

\bibitem{} B\'{e}zecourt, J., Kneib, J.-P., Soucail, G., Ebbels, T. M. D., 1999, A\&A,
  347, 21

\bibitem{} Bolton, A. S., Burles, S., Koopmans, L. V. E., Treu, T., Moustakas,
  L., A., 2006, ApJ, 638, 703

\bibitem{} Broadhurst, T., Ben\'{i}tez, N., Coe, D., Sharon, K., Zekser, K., White, R., Ford, H., et al, 2005,
ApJ, 621, 53

\bibitem{} Chary, R. -R., Stern, D., Eisenhardt, P., 2005, ApJ, 635, L5

\bibitem{} Comerford, J. M., Meneghetti, M., Bartelmann, M., Schirmer, M.,
  2006, ApJ, 642, 39

\bibitem{} Covone, G., Kneib, J. -P., Soucail, G., Richard, J., Jullo, E.,
Ebeling, H, 2006, A\&A, 456, 409

\bibitem{} Cowie, L. L., et al, 2008 in preparation

\bibitem{} Dalal, N., Holder, G., 2004, ApJ, 609, 50

\bibitem{} Ellis, R., Santos, M. R., Kneib, J. -P., Kuijken, K., 2001, ApJ,
560, L119

\bibitem{} Fedeli, C., Bartelmann, M., Meneghetti, M., Moscardini, L., 2008, A\&A submitted, preprint astro-ph/arXiv:0803.0656

\bibitem{} Frye, B., L., Coe, D., Bowen, D. V., Ben\'{i}tez, N., Broadhurst, T., Guhathakurta, P., Illingworth, G., et al, 2007,
ApJ, 665, 921

\bibitem{} Gavazzi, R., 2005, A\&A, 443, 793

\bibitem{} Gavazzi, R., Fort, B., Mellier, Y., Pello, R., Dantel-Fort, M.,
  2003, A\&A, 403, 11

\bibitem{} Grossman, S., A., Narayan, R., 1988, ApJ, 324, L37

\bibitem{} Halkola, A., Seitz, S., Pannella, M., 2006, MNRAS, 372,
1425

\bibitem{} Hamana, T., Bartelmann, M.,  Yoshida, N., Pfrommer, C., 2005,
MNRAS, 357, 1407

\bibitem{} Hammer, F., 1991, ApJ, 383, 66

\bibitem{} Hu, E. M., Cowie, L. L., McMahon, R. G., Capak, P., et
al, 2002, ApJ, 568, 75

\bibitem{}  Kneib, J. -P., Ellis, R. S., Santos, M. R., Richard, J., 2004,
  ApJ, 607, 697

\bibitem{} Kneib, J. -P., Hudelot, P., Ellis, R., Treu, T., Smith, G. P., Marshall, P., et al, 2003, ApJ, 598, 804

\bibitem{} kneib, J. -P., Mathez, G., Fort, B., Mellier, Y., Soucail, G., Longaretti,
P.-Y., 1994, A\&A, 286, 701

\bibitem{} Kneib, J.P., Mellier, Y., Fort, B., Mathez, G. 1993, A\&A, 273, 367

\bibitem{} Kovner, I., 1989, ApJ, 337, 621

\bibitem{} Leonard, A., Goldberg, D. M., Haaga, J. L., Massey, R.,
2007, ApJ, 666, L51

\bibitem{} Li, G. -L., Mao, S., Jing, Y. P., Bartelmann, M., Kang, X.,
  Meneghetti, M., 2005, ApJ, 635, 795

\bibitem{} Limousin, M., Richard, J., Jullo, E., Kneib, J. -P., Fort, B.,
Soucail, G., et al, 2006, preprint(astro-ph/0612165)

\bibitem{} Mellier, Y., Fort, B., Kneib, J. -P., 1993, ApJ, 407, 33

\bibitem{} Mellier, Y., Fort, B., Soucail, G., Mathez, G., Cailloux, M., 1991,
  ApJ, 380, 334

\bibitem{} Meneghetti, M., Jain, B., Bartelmann, M., Dolag, K., 2005a, MNRAS,
  362, 1301

\bibitem{} Meneghetti, M., Bartelmann, M., Dolag, K., Moscardini, L.,
Perrotta, F., Baccigalupi, C., Tormen, G., 2005b, A\&A, 442, 413

\bibitem{} Miradal-Escud$\acute{e}$, J., 1995, ApJ, 438, 514

\bibitem{} Miradal-Escud\'e, J., 2002, ApJ, 564, 60

\bibitem{} Navarro, J. F., Frenk, C. S., White, S. D. M., 1996, ApJ, 462, 563
(NFW)

\bibitem{} Navarro, J. F., Frenk, C. S., White, S. D. M., 1997, ApJ, 490, 493
(NFW)

\bibitem{} Ofek, E. O., Maoz, Dan.,  Rix, Hans-Walter, Kochanek, C. S.; Falco,
  E. E., 2006, ApJ, 641, 70

\bibitem{} Puchwein, E., Bartelmann, M., Dolag, K., Meneghetti, M., 2005, A\&A,
  442, 405

\bibitem{} Schimd, C., Tereno, I., Uzan, J. -P., Mellier, Y., van Waerbeke,
  L., Semboloni, E., Hoekstra, H., Fu, L., Riazuelo, A., 2006,
  preprint(astro-ph/0603158)

\bibitem{} Smith, G. P., Kneib, J. -P., Smail, I., Mazzotta, P., Ebeling, H.,
  Czoske, O., 2005, MNRAS, 359, 417

\bibitem{} Smail, I., Dressler, A., Kneib, J. -P., Ellis, R., Couch, W. J., Sharples, R. M., Oemler, A.
Jr., 1996, ApJ, 469, 508

\bibitem{} Soucail, G., Fort, B., Mellier, Y., Picat, J. P., 1987, A\&A, 187,
  L1
\bibitem{} Soucail, G., Mellier, Y., Fort, B., Cailloux, M., 1988
A\&AS, 73, 471

\bibitem{} Tu, H., Linmonsin, M., Fort, B., Shu, C., et al, 2008,
MNRAS accepted (astro-ph/0710.2246)

\bibitem{} Wambsganss, J., Bode, P., Ostriker, J.P. 2004, ApJ, 606, L93

\bibitem{} Zekser, K. C., White, R. L., Broadhurst, T. J., Ben\'{i}tez, N., Ford, H. C., Illingworth, G. D., et al., 2006, ApJ, 640, 639

\end{thebibliography}
\end{document}